\documentclass[prd,aps,showpacs,notitlepage,superscriptaddress]{revtex4-1}
\usepackage{epsfig}
\newcommand{\ben}{\begin{eqnarray}}
\newcommand{\een}{\end{eqnarray}}
\newcommand{\nnu}{\nonumber\\}
\newcommand{\bef}{\begin{figure}[htb]\centering}
\newcommand{\eef}{\end{figure}}

\begin{document}
\title{Dihadron momentum imbalance and correlations in d+Au collisions}

\date{\today}

\author{Zhong-Bo Kang}
\email{zkang@bnl.gov}
\affiliation{RIKEN BNL Research Center,
                Brookhaven National Laboratory,
                Upton, NY 11973, USA}

\author{Ivan Vitev}
\email{ivitev@lanl.gov}
\affiliation{ Theoretical Division, 
                Los Alamos National Laboratory, 
                   Los Alamos, NM 87545, USA}         
                   
\author{Hongxi Xing}
\email{xinghx@iopp.ccnu.edu.cn}
\affiliation{Institute of Particle Physics, 
                   Central China Normal University, 
                   Wuhan 430079, China}

\affiliation{Nuclear Science Division, 
                   Lawrence Berkeley National Laboratory, 
                   Berkeley, CA 94720, USA}

\begin{abstract}
We calculate in perturbative QCD the transverse momentum imbalance of dijet and dihadron 
production in  high energy p+A  (d+A)  collisions. We evaluate the effect of both 
initial- and final-state  multiple scattering, which determines the strength 
of this transverse momentum imbalance. Combining this new result with the suppression 
of the cross section in d+Au collisions, 
which arises  from cold nuclear matter energy loss and coherent power corrections,  we are able to
describe  the dihadron correlations  measured by both PHENIX and STAR collaborations at RHIC, 
including mid-mid, mid-forward, and forward-forward rapidity hadron pairs.
\end{abstract}


\maketitle

%%%%%%%%%%%%%%%%%%%%%%%%%%%%%

\section{Introduction}

In recent years, there has been renewed interest in nuclear effects that affect hadronic 
observables in proton-nucleus (p+A) and deuteron-nucleus (d+A) collisions~\cite{Boer:2011fh}. 
Much of this interest was sparked by forward rapidity measurements in d+Au collisions at the Relativistic 
Heavy Ion Collider (RHIC) that indicate considerable suppression of both single hadron  and 
dihadron production cross sections relative to the binary collision scaled proton-proton (p+p) 
baseline~\cite{Arsene:2004ux,Adler:2004eh,Adler:2005ad,Adare:2011sc,Braidot:2010ig}. 
Back-to-back dihadron correlation measurements have attracted the most attention and are 
under active investigation in the framework of different theoretical 
formalisms~\cite{Qiu:2004da,Albacete:2010pg,Strikman:2010bg,Stasto:2011ru,JalilianMarian:2011dt}. 
On one hand, quantifying the differences between p+A (d+A) and p+p collisions (referred to as 
``cold nuclear matter effects'') can provide a solid baseline 
for the unambiguous identification of any additional final-state hot and dense 
medium effects in heavy ion collisions~\cite{Gyulassy:2000er}. On the other hand, 
the study of p+A (d+A) collisions at forward rapidity is interesting in its own right
since it probes the inelastic and coherent multiple parton scattering in the nucleus, as well as
any possible modification of the small-$x$ parton distribution 
functions~\cite{Qiu:2004da,Albacete:2010pg,Strikman:2010bg,Stasto:2011ru,Marquet:2007vb,Vitev:2006bi,Kang:2008us,Kang:2007nz,JalilianMarian:2011dt}.

In perturbative QCD, at leading order, high transverse momentum ($p_\perp$) 
particle production arises from hard $2\to 2$ scattering processes. In this picture, 
two incoming collinear partons scatter on each other to produce two high transverse 
momentum partons, which then fragment into jets of hadrons. These jets, as well as the
hadron pair formed by their leading particles, are approximately back-to-back in the 
transverse plane. In high energy proton-nucleus reactions the incoming partons that 
participate in the collision can undergo multiple interactions. Even in the presence 
of multiple parton scattering,  particle production at high transverse 
momentum is dominated by a single hard interaction $Q^2 \propto p_\perp^2 \gg \xi^2$, where
$\xi^2$ is the typical scale of the soft transverse momentum exchanges. We refer to the
interactions that precede the large $Q^2$ scattering as initial-state  and the
interactions that follow the large $Q^2$ scattering as final-state.

Both initial- and final-state multiple interactions can affect  dijet (dihadron) 
production in p+A (d+A) reactions.  More specifically, they lead to an increase in the 
transverse momentum imbalance~\cite{Luo:1993ui} that can be perturbatively computed 
within a high-twist formalism~\cite{Kang:2008us, Luo:1993ui, Fries:2002mu}.  
In this paper we give a theoretical derivation of the increase in the transverse momentum
imbalance and demonstrate that it will lead to a broader away-side peak for the dihadron azimuthal 
correlation distribution in d+A collisions, consistent with the experimental observations at 
RHIC. Furthermore, multiple parton scattering also manifests itself trough medium-induced 
radiative corrections~\cite{Vitev:2007ve,Ovanesyan:2011kn} and power suppressed 
contribution to the cross sections. In reactions with nuclei (p+A and A+A) initial-state 
radiative energy loss effects are always present but their effect is most pronounced 
at forward rapidity~\cite{Vitev:2006bi,Neufeld:2010dz}. In the coherent scattering  
regime, the high-twist contributions are enhanced by the nuclear-size $\propto A^{1/3}$ and 
may become important at small and moderate transverse momenta. Such nuclear-size enhanced 
power corrections have been resummed for both inclusive single hadron and dihadron production 
processes in p+A collisions~\cite{Qiu:2004da}. In the phenomenological part of this
paper we combine the broadening in the away-side width with the nuclear cross section 
suppression to study the modification of the production rate and observed shape 
of dihadron azimuthal correlations in going from p+p to p+A (d+A) collisions. 
With parton scattering parameters  in cold nuclear matter constrained by deep inelastic scattering 
experiments~\cite{Qiu:2003vd}, calculations are consistent with the experimental 
measurements at RHIC~\cite{Braidot:2010ig,Adare:2011sc}.

The rest of our paper is organized as follows: in Sec. II we take into account both  
initial-  and final-state multiple parton scattering  to calculate the 
increase in the transverse momentum imbalance of dijet and dihadron production in p+A (d+A) 
collisions. In Sec. III we first evaluate the width of the away-side peak of the dihadron  
correlation distribution at RHIC and show that our formalism can describe the PHENIX experimental data 
well if we take into account the nuclear-induced increase in the transverse momentum imbalance. 
We then overview the nuclear effects that lead to the suppression of single and double inclusive
hadron production in p+A (d+A) reactions. We demonstrate that the calculated  
mid-forward and forward-forward rapidity hadron pair production attenuation is  
consistent with recent PHENIX measurements. At the end of this section, by combining the 
nuclear suppression and the increase in the transverse momentum imbalance, we are able to
describe the dihadron azimuthal correlation distribution observed by the STAR experiment at 
RHIC.  Summary and conclusions are presented in Sec. IV.

%%%%%%%%%%%%%%%%%%%%%%%%%%%%%
\section{Dijet and dihadron transverse momentum imbalance in $p+A$ collisions}
In this section we derive the increase in the transverse momentum imbalance of back-to-back 
jet production in p+A collisions. This phenomenon is often referred to as nuclear-induced broadening 
in the literature. We then generalize the formalism to study back-to-back hadron production.

\subsection{Dijet transverse momentum imbalance}
We start by describing the pQCD formalism for evaluating dijet production in p+p collisions:
\ben
p(P') + p(P) \to J_1(P_{1\perp})+J_2(P_{2\perp})+X.
\een
Let us define $P_\perp=|\vec{P}_{1\perp}-\vec{P}_{2\perp}|/2$ to be the magnitude of the 
average transverse momentum of the jet 
pair. In leading order perturbative QCD, the jets are produced back-to-back, $\vec{P}_{1\perp}=-\vec{P}_{2\perp}$, 
thus $|\vec{P}_{1\perp}|=|\vec{P}_{2\perp}|=P_\perp$. The differential cross section can be written 
as~\cite{Owens:1986mp}:
\ben
\frac{d\sigma}{dy_1 dy_2 dP^2_\perp}=\frac{\pi\alpha_s^2}{s^2}\sum_{a,b}
\frac{f_{a/p}(x') f_{b/p}(x)}{x' x} H^{U}_{ab\to cd}(\hat s, \hat t, \hat u) \, ,
\label{main}
\een
where $\sum_{a, b}$ runs over all parton flavors, $s=(P'+P)^2$, $f_{a, b/p}(x)$ are the parton distribution 
functions with the momentum fractions $x'=\frac{P_\perp}{\sqrt{s}}\left(e^{y_1}+e^{y_2}\right), 
x=\frac{P_\perp}{\sqrt{s}}\left(e^{-y_1}+e^{-y_2}\right)$, and $y_1$, $y_2$ the rapidites of the two jets. 
$H^{U}_{ab\to cd}(\hat s, \hat t, \hat u)$ are the partonic cross sections as a function of the usual partonic 
Mandelstam variables $\hat s, \hat t, \hat u$. These cross sections are well-known~\cite{Owens:1986mp}, 
we will reproduce them here for later convenience:
\ben
H^U_{qq'\to qq'}&=&\frac{N_c^2-1}{2N_c^2}\left[\frac{{\hat s}^2+{\hat u}^2}{{\hat t}^2}\right],
\\
H^U_{qq\to qq}&=&\frac{N_c^2-1}{2N_c^2}\left[\frac{{\hat s}^2+{\hat u}^2}{{\hat t}^2}
+\frac{{\hat s}^2+{\hat t}^2}{{\hat u}^2}\right]-\frac{N_c^2-1}{N_c^3}\left[\frac{{\hat s}^2}{{\hat t}{\hat u}}\right],
\\
H^U_{q\bar q\to q'\bar q'}&=&\frac{N_c^2-1}{2N_c^2}\left[\frac{{\hat t}^2+{\hat u}^2}{{\hat s}^2}\right],
\\
H^U_{q\bar q\to q\bar q}&=&\frac{N_c^2-1}{2N_c^2}\left[\frac{{\hat t}^2+{\hat u}^2}{{\hat s}^2}
+\frac{{\hat s}^2+{\hat u}^2}{{\hat t}^2}\right]-\frac{N_c^2-1}{N_c^3}\left[\frac{{\hat u}^2}{{\hat s}{\hat t}}\right],
\\
H^U_{qg\to qg}&=&-\frac{N_c^2-1}{2N_c^2}\left[\frac{{\hat s}}{{\hat u}}
+\frac{{\hat u}}{{\hat s}}\right]+\left[\frac{{\hat s}^2+{\hat u}^2}{{\hat t}^2}\right],
\\
H^U_{q\bar q\to gg}&=&\frac{(N_c^2-1)^2}{2N_c^3}\left[\frac{{\hat t}}{{\hat u}}
+\frac{{\hat u}}{{\hat t}}\right]-\frac{N_c^2-1}{N_c}\left[\frac{{\hat t}^2+{\hat u}^2}{{\hat s}^2}\right],
\\
H^U_{gg\to q\bar q}&=&\frac{1}{2N_c}\left[\frac{{\hat t}}{{\hat u}}
+\frac{{\hat u}}{{\hat t}}\right]-\frac{N_c}{N_c^2-1}\left[\frac{{\hat t}^2+{\hat u}^2}{{\hat s}^2}\right],
\\
H^U_{gg\to gg}&=&\frac{4N_c^2}{N_c^2-1}\left[3-\frac{{\hat t}{\hat u}}{{\hat s}^2}
-\frac{{\hat s}{\hat u}}{{\hat t}^2}-\frac{{\hat s}{\hat t}}{{\hat u}^2}\right],
\een
where $N_c=3$ is the number of colors.
\bef
\psfig{file=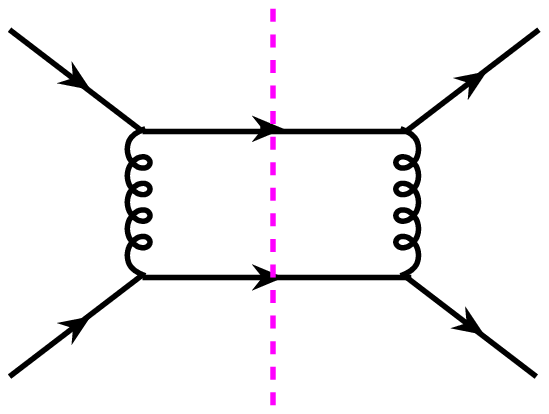, width=1.3in}
\hskip 0.3in
\psfig{file=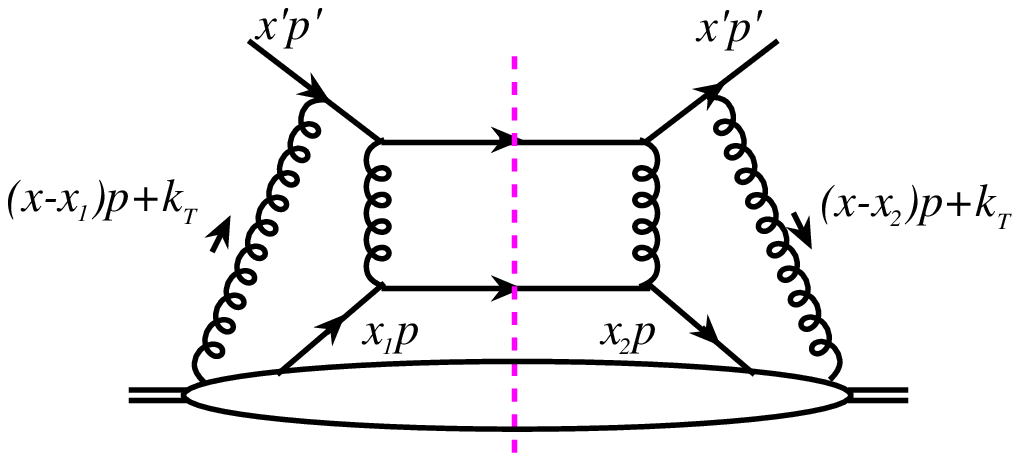, width=2.4in}
\caption{Left: Feynman diagram for the partonic  $qq'\to qq'$ scattering channel. Right:  
initial-state double scattering Feynman diagram that contributes to the dijet 
transverse momentum imbalance increase.}
\label{unini}
\eef

Let us now study dijet production in proton-nucleus collisions. In such collisions 
the energetic incoming parton from the proton can undergo multiple scattering with the 
soft partons inside the nuclear matter before the hard collisions. We refer to these  
interactions as  initial-state multiple scattering. After the hard collisions, the 
two leading partons, produced  almost back-to-back, will eventually hadronize into jets 
as observed  in the experiments. However, these partons will also likely undergo  
multiple interactions in the large nucleus. We refer to these interactions as  
final-state multiple scattering. Such separation is only possible when the
scale of hard scattering is considerably larger than the scale of typical transverse
 momentum exchanges between the projectile and the medium, $Q^2 \gg \xi^2$~\cite{Neufeld:2010dz}.
The initial- and final-state multiple scattering lead to dijet acoplanarity, or imbalance. 
To quantify this effect, let us define the dijet transverse 
momentum imbalance $\vec{q}_\perp$ as:
\ben
\vec{q}_\perp=\vec{P}_{1\perp}+\vec{P}_{2\perp},
\een
and  the averaged transverse momentum imbalance square as:
\ben
\langle q_\perp^2\rangle = \left. \int d^2 \vec{q}_\perp q_\perp^2 
\frac{d\sigma}{dy_1 dy_2 dP^2_\perp d^2\vec{q}_\perp}\right/ \frac{d\sigma}{dy_1 dy_2 dP^2_\perp} .
\een
The nuclear enhancement (or broadening) of the transverse momentum imbalance in p+A 
collisions relative to p+p collisions can be quantified by the difference:
\ben
\Delta \langle q_\perp^2\rangle = \langle q_\perp^2\rangle_{pA} - \langle q_\perp^2\rangle_{pp}.
\een

As demonstrated in Ref.~\cite{Kang:2008us}, the transverse momentum imbalance increase 
$\Delta \langle q_\perp^2\rangle$ can be calculated perturbatively and the leading contribution 
comes from double scattering. To illustrate the method, we will study 
the simple partonic channel $qq'\to qq'$. At lowest order, the partonic cross section 
$H^U_{qq'\to qq'}$ is calculated from the Feynman diagram in Fig.~\ref{unini} (left). At this order, 
the transverse momenta of the two outgoing partons (or the dijet) are equal and opposite and we have:
\ben
\frac{d\sigma}{dy_1 dy_2 dP^2_\perp d^2\vec{q}_\perp} \propto \delta^2(\vec{q}_\perp).
\een
The leading contribution to the transverse momentum imbalance increase in p+A collisions comes 
from either  initial-state double scattering, as in Fig.~\ref{unini} (right); 
or  final-state double scattering, as in Fig.~\ref{fin}. 
Let us first focus on the calculation of initial-state double scattering. In this case, 
the small $k_\perp$-kick  in the nucleus will generate a small transverse momentum 
imbalance for the dijet:
\ben
\vec{q}_\perp = \vec{P}_{1\perp} + \vec{P}_{2\perp} = \vec{k}_\perp.
\een
Following~\cite{Kang:2008us}, we evaluate the contribution from the double 
scattering diagram to $\langle q^2_\perp\rangle_{pA}$:
\ben
\int d^2\vec{q}_\perp q_\perp^2\, \frac{d\sigma_{pA}}{dy_1 dy_2 dP^2_\perp d^2\vec{q}_\perp}
&=&
\frac{\pi\alpha_s^2}{s^2} \frac{f_{q/p}(x')}{x' x} \int d^2\vec{q}_\perp q_\perp^2 \int dx_1 dx_2
d^2k_\perp \overline{T}_{Aq}(x, x_1, x_2, k_\perp)
\nnu
&&\times
\overline{H}(x,x_1,x_2,k_\perp,x'p') \delta^2(\vec{q}_\perp-\vec{k}_\perp)\, ,
\een
where the matrix element $\overline{T}_{Aq}$ is represented by the bottom blob in 
Fig.~\ref{unini} (right)  and has the following expression:
\ben
\overline{T}_{Aq}(x,x_1,x_2,k_\perp)
=
\int\frac{dy^-}{2\pi}\, \frac{dy_1^-}{2\pi}\,
\frac{dy_2^-}{2\pi}\,
\int\frac{d^2y_\perp}{(2\pi)^2}\, 
e^{ix_1p^+y_1^-}e^{i(x-x_1)p^+y^-}e^{-i(x-x_2)p^+y_2^-}
e^{-ik_\perp\cdot y_\perp}
\nonumber\\ 
\times 
\frac{1}{2}\langle p_A|
A^+(y_2^-,0_\perp)\bar{\psi}(0)\gamma^+\psi(y_1^-)A^+(y^-,y_\perp)
|p_A\rangle.
\label{Tbar}
\een
As in~\cite{Kang:2008us, Luo:1993ui}, the calculation first  takes  advantage of 
$\delta^2(\vec{q}_\perp-\vec{k}_\perp)$ to set $q_\perp^2=k_\perp^2$. These two factors of $k_\perp$ 
can be converted into transverse derivatives on the two fields $A^+$ in matrix element $\overline{T}_{Aq}$ 
in Eq.~(\ref{Tbar}) by performing a partial integration for $d^2y_\perp$:
\ben
k_\perp^2\,A^+(y_2^-,0_\perp)A^+(y^-,y_\perp) 
\to
F_\alpha^{~+}(y_2^-,0_\perp)F^{+\alpha}(y^-,y_\perp).
\een
We then expand in $k_\perp$ in the partonic part $\overline{H}$ around $k_\perp^2=0$, and keep 
the first non-vanishing term: 
\ben
H(x,x_1,x_2,x'p')\equiv\overline{H}(x,x_1,x_2,k_\perp=0,x'p').
\een
In other words, since the $A^+$ fields have been converted to the gauge-covariant gluon field 
strength in the matrix element $\overline{T}_{Aq}$, one can set $k_\perp=0$ in the hard part. 
One will later find that all such hard-parts $H(x,x_1,x_2,x'p')$ for the double scattering diagrams 
reduce to the lowest order Born diagrams up to color factors. Thus, the calculation of transverse 
momentum imbalance enhancement $\Delta\langle q_\perp^2\rangle$ is simpler than the usual 
higher-twist calculations in which one calculates the multiple scattering contribution to the 
differential cross section itself directly (instead of the $q_\perp^2$-weighted one).

We obtain:
\ben
\int d^2\vec{q}_\perp q_\perp^2\, \frac{d\sigma_{pA}}{dy_1 dy_2 dP^2_\perp d^2\vec{q}_\perp}
=\frac{\pi\alpha_s^2}{s^2} \frac{f_{q/p}(x')}{x' x} \int dx_1 dx_2
T_{Fq}(x, x_1, x_2) H(x,x_1,x_2,x'p'),
\label{qt2}
\een
where $T_{Fq}$ is a twist-4 four-parton correlation function:
\ben
T_{Fq}^{(I)}(x,x_1,x_2)=
\int\frac{dy^-}{2\pi}\, \frac{dy_1^-}{2\pi}\, \frac{dy_2^-}{2\pi}\, 
e^{ix_1p^+y_1^-}e^{i(x-x_1)p^+y^-}e^{-i(x-x_2)p^+y_2^-}
\frac{1}{2}\langle p_A|
F_\alpha^{~+}(y_2^-)\bar{\psi}(0)\gamma^+\psi(y_1^-)F^{+\alpha}(y^-)
|p_A\rangle. \quad
\label{TF}
\een
Here and thereafter we will use the superscript ``I'' (``F'') to indicate the contribution associated 
with initial- (final-) state multiple scattering. The hard partonic function $H(x,x_1,x_2,x'p')$ 
is given by:
\ben
H(x,x_1,x_2,x'p')
=8\pi^2\alpha_s \frac{C_F}{N_c^2-1}
\left[\frac{1}{2\pi}\frac{1}{x_1-x-i\epsilon}
\frac{1}{x_2-x+i\epsilon}\right] H_{qq'\to qq'}^U,
\label{H}
\een
where $C_F=(N_c^2-1)/2N_c$. Substituting Eq.~(\ref{H}) into Eq.~(\ref{qt2}) and performing the 
integration over $x_1$ and $x_2$, we obtain: 
\ben
\int d^2\vec{q}_\perp q_\perp^2\, \frac{d\sigma_{pA}}{dy_1 dy_2 dP^2_\perp d^2\vec{q}_\perp}
=\left(\frac{8\pi^2\alpha_s}{N_c^2-1}\right)\frac{\pi\alpha_s^2}{s^2} 
\frac{f_{q/p}(x')}{x' x} T_{q/A}^{(I)}(x) H^I_{qq'\to qq'} (\hat s, \hat t, \hat u),
\een
where $T_{q/A}^{(I)}(x)$ is the twist-4 quark-gluon correlation function defined 
as~\cite{Kang:2008us, Luo:1993ui}:
\ben
T_{q/A}^{(I)}(x) =
 \int \frac{dy^{-}}{2\pi}\, e^{ixp^{+}y^{-}}
 \int \frac{dy_1^{-}dy_{2}^{-}}{2\pi} \,
      \theta(y^{-}-y_{1}^{-})\,\theta(-y_{2}^{-}) 
     \frac{1}{2}\,
     \langle p_{A}|F_{\alpha}^{\ +}(y_{2}^{-})\bar{\psi}_{q}(0)
                  \gamma^{+}\psi_{q}(y^{-})F^{+\alpha}(y_{1}^{-})
     |p_{A} \rangle,
\label{TqA}
\een
and $H^I_{qq'\to qq'}$ is the hard-part function given by:
\ben
H^I_{qq'\to qq'}=C_F H^U_{qq'\to qq'}.
\een
\bef
\psfig{file=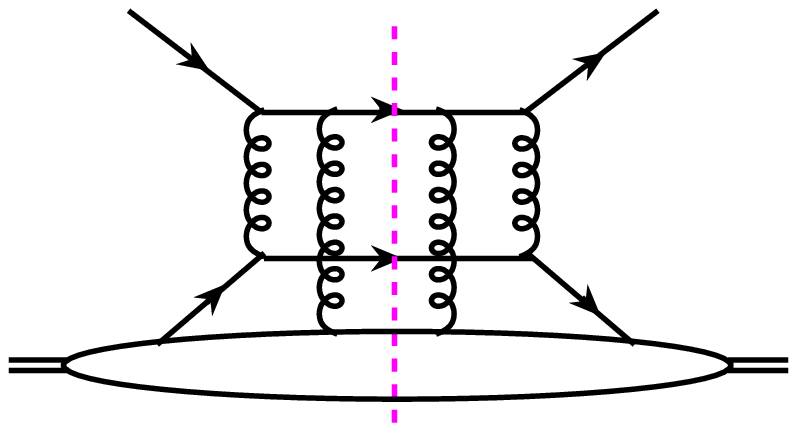, width=1.5in}
\hskip 0.1in
\psfig{file=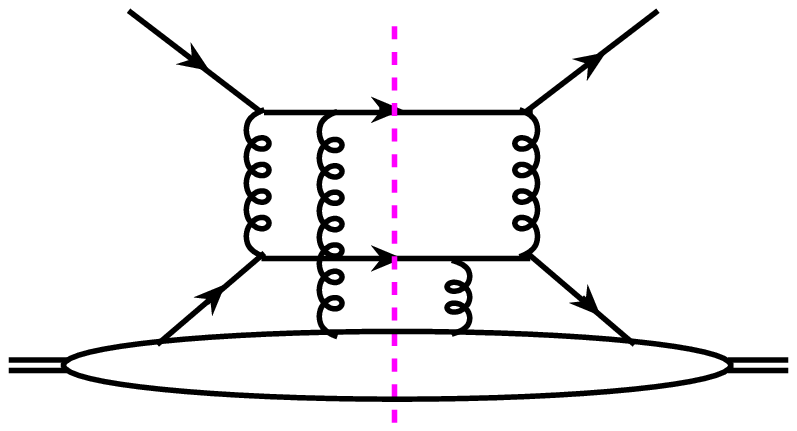, width=1.5in}
\hskip 0.1in
\psfig{file=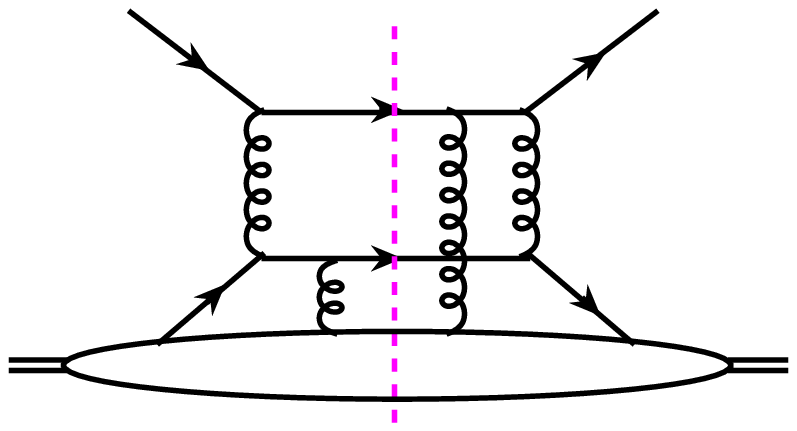, width=1.5in}
\hskip 0.1in
\psfig{file=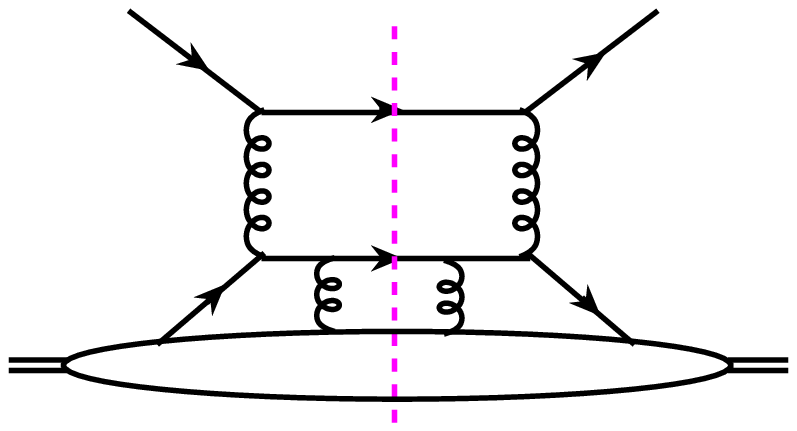, width=1.5in}
\caption{Final-state double scattering Feynman diagrams that contribute to the dijet 
transverse momentum imbalance increase, for the partonic channel $qq'\to qq'$.}
\label{fin}
\eef

Next, we calculate the contribution to the transverse momentum imbalance from the 
final-state double scattering diagrams in Fig.~\ref{fin}. This evaluation is slightly more 
complicated since there are four diagrams as opposed to a single diagram for the initial-state case. 
However, the calculation is very similar and straightforward. Different Feynman diagrams 
only differ  in terms of color factors  and their sum provides a measure of the 
average square transverse field strength probed by the outgoing partons~\cite{Luo:1993ui}. 
We obtain:
\ben
\int d^2\vec{q}_\perp q_\perp^2\, \frac{d\sigma_{pA}}{dy_1 dy_2 dP^2_\perp d^2\vec{q}_\perp}
=\left(\frac{8\pi^2\alpha_s}{N_c^2-1}\right)\frac{\pi\alpha_s^2}{s^2} 
\frac{f_{q/p}(x')}{x' x} T_{q/A}^{(F)}(x) H^F_{qq'\to qq'} (\hat s, \hat t, \hat u),
\een
where $T_{q/A}^{(F)}$ is given by the same expressions in Eq.~(\ref{TqA}), 
except for the $\theta$-functions that are replaced as follows~\cite{Kang:2008us, Luo:1993ui}:
\ben
\theta(y^{-}-y_{1}^{-})\,\theta(-y_{2}^{-}) 
\to
\theta(y_{1}^{-}-y^{-})\,\theta(y_{2}^{-}),
\label{theta}
\een
and the final-state hard-part function $H^F$ is given by
\ben
H^F_{qq'\to qq'}=\frac{(N_c^2-3)(N_c^2-1)}{2N_c^3}\left[\frac{{\hat s}^2+{\hat u}^2}{{\hat t}^2}\right].
\een

Likewise, we can go ahead to calculate the contribution to the transverse momentum imbalance from both 
initial- and final-state double scattering diagrams for all  partonic channels. 
The calculation is straightforward though tedious, the result can be summarized as:
\ben
\int d^2\vec{q}_\perp q_\perp^2\, \frac{d\sigma_{pA}}{dy_1 dy_2 dP^2_\perp d^2\vec{q}_\perp} =
\left(\frac{8\pi^2\alpha_s}{N_c^2-1}\right)
\frac{\pi\alpha_s^2}{s^2} \sum_{a, b}\frac{f_{a/p}(x')}{x' x} 
\left[T_{b/A}^{(I)}(x) H^I_{ab\to cd} (\hat s, \hat t, \hat u)+
T_{b/A}^{(F)}(x) H^F_{ab\to cd} (\hat s, \hat t, \hat u)\right],
\label{numerator}
\een
where $T_{q/A}^{(I, F)}(x)$ are given in Eqs.~(\ref{TqA}) and (\ref{theta}). 
In the calculation for the partonic channels $qg\to qg$ and $gg\to gg$, see for example 
the Feynman diagrams shown in Fig.~\ref{gg}, two other similar twist-4 gluon-gluon correlation functions 
$T_{g/A}^{(I, F)}(x)$ appear. The operator definition of $T_{g/A}^{(I)}(x)$ is given 
by~\cite{Luo:1993ui, Kang:2008us}:
\ben
T_{g/A}^{(I)}(x) &=&
 \int \frac{dy^{-}}{2\pi}\, e^{ixp^{+}y^{-}}
 \int \frac{dy_1^{-}dy_{2}^{-}}{2\pi} \,
      \theta(y^{-}-y_{1}^{-})\,\theta(-y_{2}^{-}) 
\frac{1}{xp^+}\,
\langle p_A| F_\alpha^{~+}(y_2^-)
F^{\sigma+}(0)F^+_{~\sigma}(y^-)F^{+\alpha}(y_1^-)|p_A\rangle\, ,
\label{TgA}
\een
while $T_{g/A}^{(F)}(x)$ is given by the same expression with the $\theta$-function 
replacement specified in Eq.~(\ref{theta}).
\bef
\psfig{file=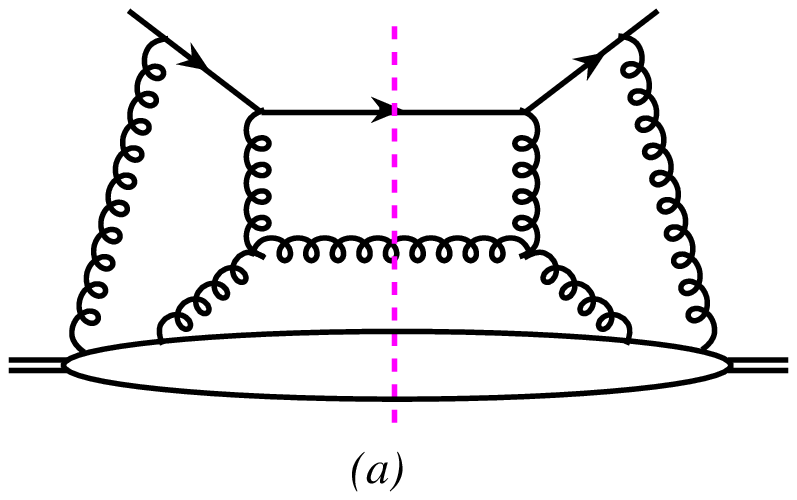, width=1.5in}
\hskip 0.1in
\psfig{file=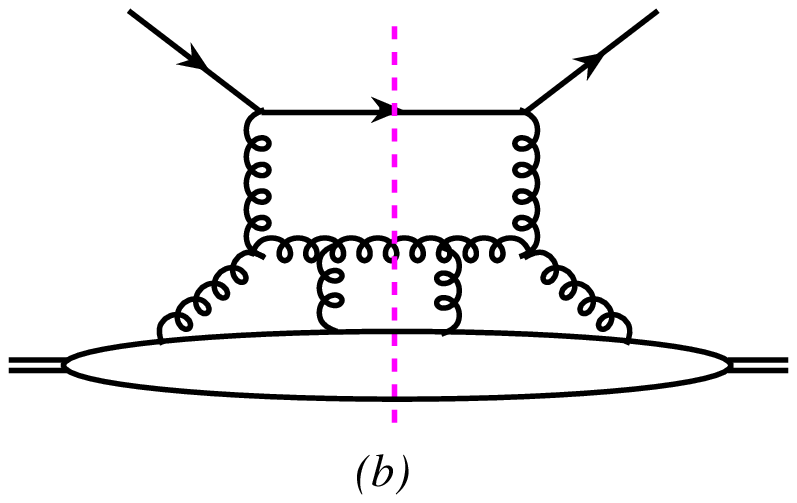, width=1.5in}
\hskip 0.1in
\psfig{file=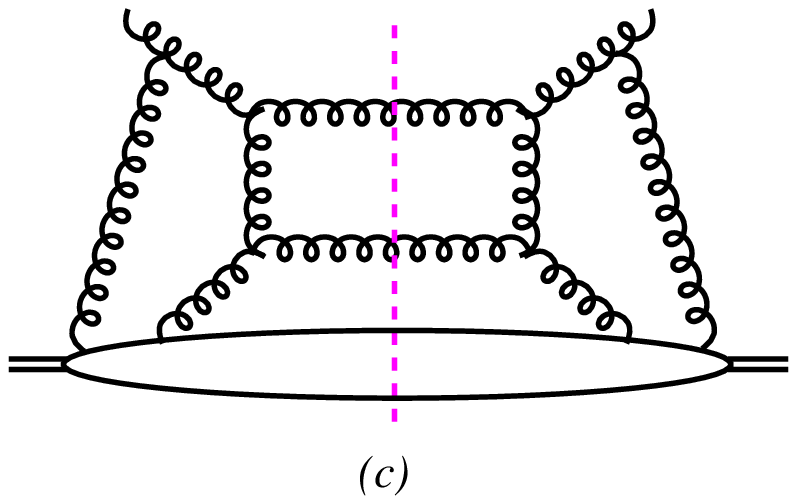, width=1.5in}
\hskip 0.1in
\psfig{file=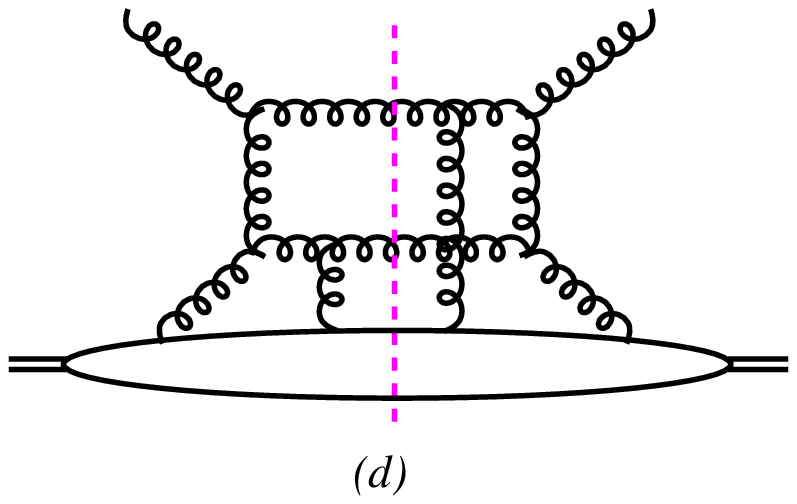, width=1.5in}
\caption{Sample diagrams that contribute to the dijet transverse momentum 
imbalance for the partonic channels $qg\to qg$ (left two) and $gg\to gg$ (right two), 
where (a) and (c) are for initial-state double scattering, (b) and (d) are for 
final-state double scattering.}
\label{gg}
\eef

Finally, from Eqs.~(\ref{main}) and (\ref{numerator}), we obtain the transverse momentum imbalance 
increase $\Delta\langle q_\perp^2\rangle$ in $pA$ collisions:
\ben
\Delta\langle q_\perp^2\rangle=\left(\frac{8\pi^2\alpha_s}{N_c^2-1}\right)
\frac{ \sum_{a, b}\frac{f_{a/p}(x')}{x' x} 
\left[T_{b/A}^{(I)}(x) H^I_{ab\to cd} (\hat s, \hat t, \hat u)+
T_{b/A}^{(F)}(x) H^F_{ab\to cd} (\hat s, \hat t, \hat u)\right]}
{\sum_{a,b}\frac{f_{a/p}(x') f_{b/p}(x)}{x' x} H^{U}_{ab\to cd}(\hat s, \hat t, \hat u)}.
\label{imbalance}
\een
The partonic hard-part functions $H^I_{ab\to cd}$ are associated with the initial-state multiple
 scattering, and are given by:
\ben
H^I_{ab\to cd} = \left\{
  \begin{array}{l l}
    C_F H^U_{ab\to cd} & \quad \text{a=quark}\\
     \\
    C_A H^U_{ab\to cd} & \quad \text{a=gluon}\\
  \end{array} \, , \right.  
\een
with $C_A=N_c$. On the other hand, $H^F_{ab\to cd}$ are associated with the final-state multiple scattering, 
and are given by:
\ben
H^F_{qq'\to qq'}&=&\frac{(N_c^2-3)(N_c^2-1)}{2N_c^3}\left[\frac{{\hat s}^2+{\hat u}^2}{{\hat t}^2}\right],
\\
H^F_{qq\to qq}&=&\frac{(N_c^2-3)(N_c^2-1)}{2N_c^3}\left[\frac{{\hat s}^2+{\hat u}^2}{{\hat t}^2}
+\frac{{\hat s}^2+{\hat t}^2}
{{\hat u}^2}\right]+\frac{2(N_c^2-1)}{N_c^4}\left[\frac{{\hat s}^2}{{\hat t}{\hat u}}\right],
\\
H^F_{q\bar q\to q'\bar q'}&=&C_A H^U_{q\bar q\to q'\bar q'},
\\
H^F_{q\bar q\to q\bar q}&=&C_A H^U_{q\bar q\to q\bar q}
-\frac{(N_c^2-1)^2}{2N_c^3}\left[\frac{{\hat s}^2+{\hat u}^2}{{\hat t}^2}\right],
\\
H^F_{qg\to qg}&=&C_F H^U_{qg\to qg} -\frac{N_c}{2} \left[\frac{\hat s(\hat s^2+\hat u^2)}
{\hat t^2\hat u}\right],
\\
H^F_{q\bar q \to gg}&=&C_A H^U_{q\bar q\to gg}
-\frac{N_c^2-1}{2N_c^2}\left[\frac{\hat t}{\hat u}+\frac{\hat u}{\hat t}\right],
\\
H^F_{gg\to q\bar{q}}&=&C_A H^U_{gg\to q\bar{q}}-\frac{1}{2(N_c^2-1)}
\left[\frac{\hat t}{\hat u}+\frac{\hat u}{\hat t}\right],
\\
H^F_{gg\to gg}&=&C_A H^U_{gg\to gg}+\frac{2N_c^3}{(N_c^2-1)^2}
\left[\frac{\hat t}{\hat u}+\frac{\hat u}{\hat t}+1\right]^2.
\een

\subsection{Dihadron transverse momentum imbalance}
One can easily generalize the transverse momentum imbalance for dijet production 
to dihadron production. To start, we write down the leading order differential cross 
section for dihadron production in p+p collisions, 
$p(P') + p(P) \to h_1(p_{1\perp})+h_2(p_{2\perp})+X$,
\ben
\frac{d\sigma}{dy_1 dy_2 dp_{1\perp} dp_{2\perp}}=\frac{2\pi\alpha_s^2}{s^2}\sum_{abcd}
\int \frac{dz_1}{z_1}D_{h_1/c}(z_1) D_{h_2/d}(z_2) 
\frac{f_{a/p}(x') f_{b/p}(x)}{x' x} H^{U}_{ab\to cd}(\hat s, \hat t, \hat u),
\label{dihadron}
\een
where $D_{h_1/c}(z_1)$ and $D_{h_2/d}(z_2)$ are fragmentation functions, 
 $z_2=z_1\, p_{2\perp}/p_{1\perp}$~\cite{Vitev:2006bi}, and:
\ben
x'=\frac{p_{1\perp}}{z_1\sqrt{s}}\left(e^{y_1}+e^{y_2}\right), \qquad 
x=\frac{p_{1\perp}}{z_1\sqrt{s}}\left(e^{-y_1}+e^{-y_2}\right).
\een
At this order, the final hadron pair comes from the fragmentation of the back-to-back 
parton pair in the partonic collision $ab\to cd$. One has $\vec{p}_{1\perp} = 
z_1 \vec{P}_\perp$ and $\vec{p}_{2\perp} = -z_2 \vec{P}_\perp$ with $\vec{P}_\perp$ 
the transverse momentum of the first parent parton. Then,
the dihadron transverse momentum imbalance $\vec{q}_\perp$ can be written as:
\ben
\vec{q}_\perp \equiv \vec{p}_{1\perp}+\vec{p}_{2\perp}=(z_1 - z_2) \vec{P}_{\perp}.
\een
Thus, the averaged transverse momentum imbalance $\langle q_\perp^2 \rangle$ in p+p collisions is:
\ben
\langle q_\perp^2 \rangle_{pp} = \left. \int d^2 \vec{q}_\perp q_\perp^2 
\frac{d\sigma}{dy_1 dy_2 dp_{1\perp} dp_{2\perp} d^2\vec{q}_\perp}\right 
/ \frac{d\sigma}{dy_1 dy_2 dp_{1\perp} dp_{2\perp} } 
\sim \left\langle (z_1 - z_2)^2 P_{\perp}^2 \right\rangle.
\een
On the other hand, in p+A collisions there will be transverse momentum imbalance 
increase generated by the small $k_\perp$-kick from the initial- and final-state 
multiple scattering in the nuclear medium, as demonstrated in the previous section. 
Since the two effects are independent: 
\ben
\langle q_\perp^2 \rangle_{pA} = 
\left\langle ((z_1-z_2)\vec{P}_\perp+\vec{k}_\perp)^2\right\rangle
=\left\langle (z_1 - z_2)^2 P_{\perp}^2 \right\rangle + \langle k_\perp^2\rangle,
\een
where in the second step we have taken $\langle (z_1-z_2) \vec{P}_\perp \cdot \vec{k}_\perp\rangle=0$ 
due to the random nature of the $k_\perp$-kick. Thus, the transverse momentum imbalance 
increase in p+A collisions when compared to p+p collisions is given by:
\ben
\Delta\langle q_\perp^2 \rangle=\langle q_\perp^2 \rangle_{pA} -\langle q_\perp^2 \rangle_{pp} 
=\langle k_\perp^2 \rangle,
\een
and is equal to the strength of the soft $k_\perp$-kick from the multiple scattering. 
From the result of the last section, we immediately obtain for the dihadron production:
\ben
\Delta\langle q_\perp^2 \rangle =\left(\frac{8\pi^2\alpha_s}{N_c^2-1}\right)
\frac{ \sum_{abcd}\int \frac{dz_1}{z_1}D_{h_1/c}(z_1) D_{h_2/d}(z_2) \frac{f_{a/p}(x')}{x' x} 
\left[T_{b/A}^{(I)}(x) H^I_{ab\to cd} (\hat s, \hat t, \hat u)+
T_{b/A}^{(F)}(x) H^F_{ab\to cd} (\hat s, \hat t, \hat u)\right]}
{\sum_{abcd}\int \frac{dz_1}{z_1}D_{h_1/c}(z_1) D_{h_2/d}(z_2) 
\frac{f_{a/p}(x') f_{b/p}(x)}{x' x} H^{U}_{ab\to cd}(\hat s, \hat t, \hat u)} . \quad
\label{imbalancehadron}
\een
Eqs.~(\ref{imbalance}) and (\ref{imbalancehadron}) are the main new theoretical 
results of our paper.  We will use them in the phenomenological studies of dihadron 
azimuthal correlations in the next section.

%%%%%%%%%%%%%%%%%%%

\section{Dihadron correlation in $d+Au$ collisions}

In this section, we study the phenomenological applications of our results. 
First, we  use the transverse momentum imbalance result to derive the width  of the 
away-side peak in dihadron correlations and compare our findings to existing experimental data. 
Next, we review the nuclear effects which lead to the suppression of dihadron 
production in d+Au collisions  and  calculate the nuclear modification factor $R_{dA}^{(2)}$, 
which turns out to describe the data reasonably well. Finally combining the calculation on 
both the away-side width and the nuclear suppression factor, we are able to describe 
the broadening of the dihadron azimuthal correlations in d+Au collisions relative 
to p+p collisions.

\subsection{Dihadron imbalance and the width of the away-side peak}

In order to estimate the nuclear-induced broadening for dihadron production given by 
Eq.~(\ref{imbalancehadron}), one needs to know the four-parton correlation functions 
$T_{b/A}^{(I, F)}(x)$ with $b=q,g$. Following Refs.~\cite{Qiu:2003vd, Qiu:2004da}, these can be 
modeled as:
\ben
\frac{4\pi^2\alpha_s}{N_c}\,T_{q,g/A}^{(I)}(x)=\frac{4\pi^2\alpha_s}{N_c}\,T_{q,g/A}^{(F)}(x) 
= \xi^2 \left(A^{1/3}-1\right) f_{q,g/A}(x). 
\label{ht}
\een
Such decomposition into a leading twist parton distribution function and nuclear 
enhanced transverse momentum  transfers squared and/or power corrections is motivated 
by the $\theta$-function structure, and the similarity of the operator definition between $T_{q,g/A}^{(I,F)}(x)$ and $f_{q,g/A}(x)$. In Eq.~(\ref{ht}) $\xi^2=0.09-0.12$~GeV$^2$ 
represents a characteristic scale of high twist corrections
per nucleon  and has been extracted from deep inelastic scattering data~\cite{Qiu:2003vd}. 
It has also been employed to describe  single inclusive hadron production 
in d+Au collisions at forward rapidities~\cite{Qiu:2004da,Vitev:2006bi,Neufeld:2010dz}.
We can now write the increase of transverse momentum imbalance as:
\ben
\Delta\langle q_\perp^2 \rangle =\frac{2N_c}{N_c^2-1}\xi^2\left(A^{1/3}-1\right)
\frac{ \sum_{abcd}\int \frac{dz_1}{z_1}D_{h_1/c}(z_1) D_{h_2/d}(z_2) \frac{f_{a/p}(x') f_{b/A}(x) }{x' x} 
 \left[H^I_{ab\to cd} (\hat s, \hat t, \hat u)+H^F_{ab\to cd} (\hat s, \hat t, \hat u)\right]}
{\sum_{abcd}\int \frac{dz_1}{z_1}D_{h_1/c}(z_1) D_{h_2/d}(z_2) 
\frac{f_{a/p}(x') f_{b/p}(x)}{x' x} H^{U}_{ab\to cd}(\hat s, \hat t, \hat u)}. \qquad
\label{dihadronimbalance}
\een
The transverse momentum imbalance 
$\langle q_\perp^2 \rangle_{dA}=\langle q_\perp^2 \rangle_{pp}+\Delta\langle q_\perp^2 \rangle$ 
characterizes the shape of the azimuthal correlation of back-to-back dihadron production 
in d+Au collisions, as it is closely related to the width of the away-side peak $\sigma_F$. 
First, we have the following relation~\cite{Rak:2003ay, Angelis:1980bs}:
\ben
\langle |p_{\rm out}|\rangle^2=\langle |j_{\perp y}|\rangle ^2
+x_E^2 \left(\langle |j_{\perp y}|\rangle ^2+2\langle |k_{\perp y}|\rangle^2_{\rm parton}\right) \, ,
\een
where $\langle |p_{\rm out}|\rangle$ is the average transverse momentum out of the plane 
defined by the momentum of the trigger particle $\vec{p}_{\perp, \rm trig}$ and the beam axis, 
$j_{\perp y}$ is the component of the particle momentum perpendicular to the jet momentum in the 
fragmentation process, and $\langle |k_{\perp y}|\rangle_{\rm parton}^2=
\langle k_{\perp}^2\rangle_{\rm parton}/\pi$. On the other hand, 
partonic $\langle k_{\perp}^2\rangle_{\rm parton}$ is related to the transverse momentum 
imbalance as $\langle q_{\perp}^2\rangle = 2 \langle k_{\perp}^2\rangle_{\rm parton}$. Realizing that:
\ben
\langle |p_{\rm out}|\rangle = p_{\perp, \rm assoc} \sin\left( |\Delta \phi|\right)
=p_{\perp, \rm assoc}\sin\left(\sqrt{\frac{2}{\pi}}\sigma_F\right)\, ,
\een
where $p_{\perp, \rm assoc}$ is the transverse momentum of the associated particle, and
\ben
x_E=-\frac{\vec{p}_{\perp, \rm assoc}\cdot \vec{p}_{\perp, \rm trig}}{\vec{p}_{\perp, \rm trig}^2}
\approx -\frac{p_{\perp, \rm assoc}}{p_{\perp, \rm trig}} \cos\left(|\Delta \phi|\right) 
= -\frac{p_{\perp, \rm assoc}}{p_{\perp, \rm trig}}\cos\left(\sqrt{\frac{2}{\pi}}\sigma_F\right)\, ,
\een
we obtain:
\ben
\cos^2\left(\sqrt{\frac{2}{\pi}}\sigma_F\right)=
\frac{1-\frac{\langle |j_{\perp y}|\rangle^2}{p_{\perp, \rm assoc}^2}}
{1+\frac{\langle |j_{\perp y}|\rangle^2}{p_{\perp, \rm trig}^2}
+\frac{1}{\pi}\frac{\langle q_{\perp}^2\rangle}{p_{\perp, \rm trig}^2}}\, .
\label{sigmaF}
\een
It has been shown that $\langle |j_{\perp y}|\rangle\approx 400$ MeV is sensitive 
only to vacuum fragmentation and is independent of the center-of-mass energy 
$\sqrt{s}$ and the trigger particle momentum~\cite{Rak:2003ay, Angelis:1980bs}. 
Thus, for the selected trigger and associated particles (with specific momenta), the 
width of the away-side peak $\sigma_F$ depends closely on the transverse momentum 
imbalance $\langle q_\perp^2\rangle$. 

\bef
\psfig{file=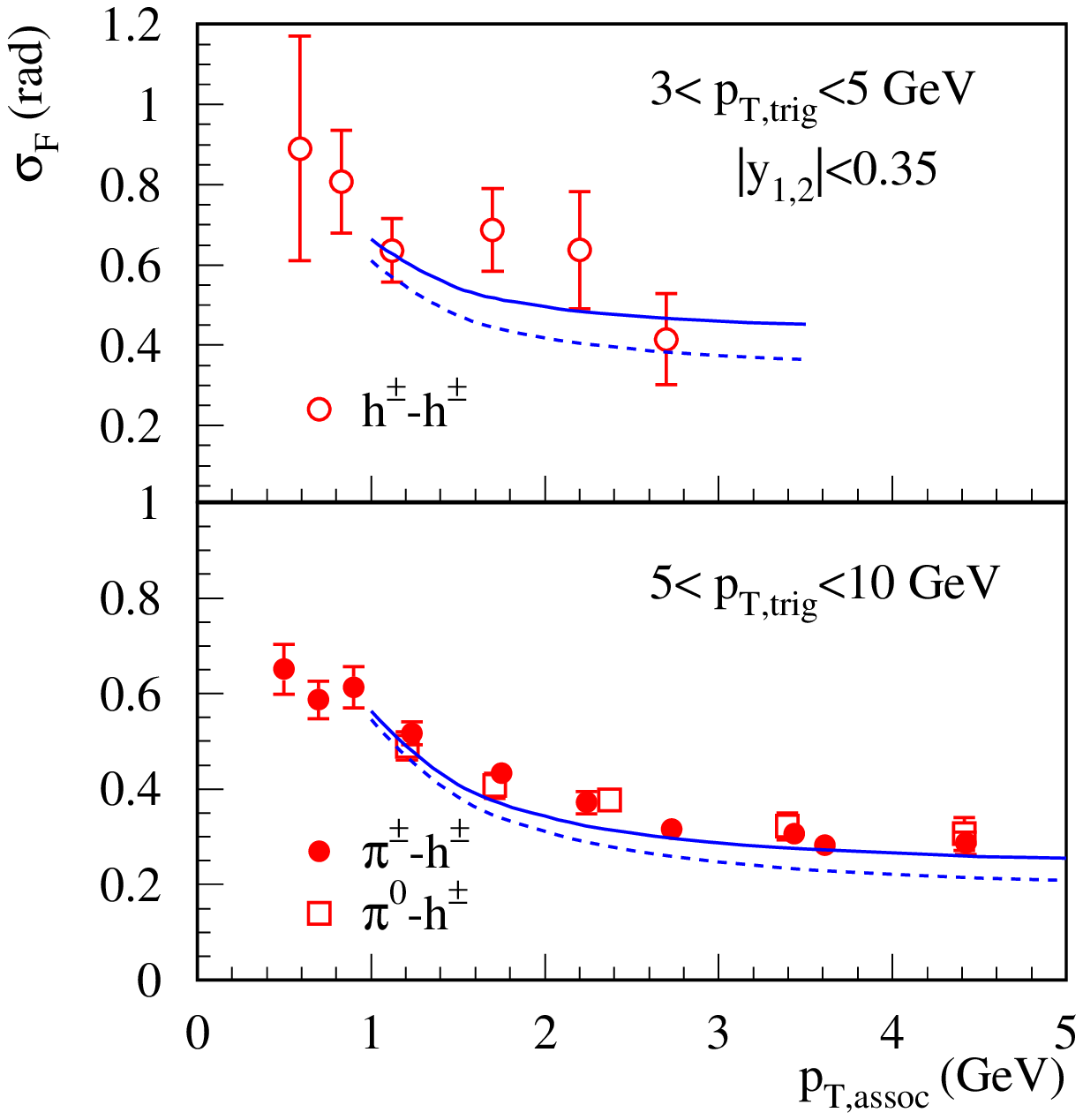, width=3.2in}
\caption{The width $\sigma_F$ of the away-side correlation is plotted as a function of 
the associated hadron transverse momentum $p_{\perp,\rm assoc}$. Both hadrons are at mid-rapidity 
$|y_{1,2}|<0.35$. The upper panel is for $3<p_{\perp,\rm trig}<5$ GeV 
and the lower panel is for $5<p_{\perp,\rm trig}<10$ GeV. Solid curves represent the 
width calculated  from $\langle q_\perp^2\rangle_{dA}=\langle q_\perp^2\rangle_{pp}+\Delta\langle q_\perp^2\rangle$ 
and include multiple scattering effect. Dashed are calculated from
$\langle q_\perp^2\rangle_{pp}$. Data is from PHENIX~\cite{Adler:2005ad}.}
\label{awaysigma}
\eef
The PHENIX collaboration at RHIC has measured $\sigma_F$, as shown in Fig.~\ref{awaysigma}. 
We first compute $\Delta\langle q_\perp^2\rangle$ by taking $\xi^2=0.12$ GeV$^2$~\cite{Qiu:2004da} 
and by using the CTEQ6L1 parton distribution functions~\cite{Pumplin:2002vw} and the 
deFlorian-Sassot-Stratmann (DSS) hadron fragmentation functions~\cite{deFlorian:2007aj}. 
We choose $\langle q_\perp^2\rangle_{pp}\approx 3.3$ GeV$^2$ extracted from PHENIX p+p data~\cite{Rak:2003ay} 
to evaluate $\langle q_\perp^2\rangle_{dA}$ in d+Au collisions.

In Fig.~\ref{awaysigma} the width $\sigma_F$ of the away-side peak is plotted as a 
function of the associated particle transverse momentum $p_{\perp,\rm assoc}$ for two ranges 
of the trigger particle momentum in the minimum bias d+Au collisions. The solid curves 
are calculated from $\langle q_\perp^2\rangle_{dA}$, while the dashed ones are 
from $\langle q_\perp^2\rangle_{pp}$ (i.e., without multiple scattering). Even though 
the increase in the transverse momentum imbalance $\Delta\langle q_\perp^2\rangle$ is 
on the order of $2-3$~GeV$^2$, the broadening effect in the width $\sigma_F$ is moderate 
for the mid-mid correlation, where both trigger and associated particles are in 
the mid-rapidity region $|y_{1,2}|<0.35$. Nevertheless, the calculated $\sigma_F$ 
with multiple scattering effects appears to agree with the PHENIX data slightly 
better.

\subsection{Nuclear modification factor in d+Au collisions}
Coherent multiple parton scattering can also affect the rate of dihadron production. 
The difference between p+A (d+A) and p+p reactions  is usually quantified by the nuclear 
modification factor:
\ben
R_{dA}^{(2)} = \frac{d\sigma_{dA}/dy_1 dy_2 dp_{1\perp} dp_{2\perp}}
{\langle N_{coll} \rangle d\sigma_{pp}/dy_1 dy_2 dp_{1\perp} dp_{2\perp}} \, .
\een
There are two major effects that control the magnitude of $R_{dA}^{(2)}$:  dynamical 
shadowing~\cite{Qiu:2004da, Vitev:2006bi} 
and cold nuclear matter energy loss~\cite{Vitev:2007ve,Neufeld:2010dz,Xing:2011fb}. 
Dynamical shadowing effects have been shown in Refs.~\cite{Qiu:2004da, Vitev:2006bi} 
to contribute to the cross section at the power corrections level. Nuclear size 
enhanced ($A^{1/3}$) power corrections can be resummed for a given partonic 
channel, as shown by Qiu and Vitev, and lead to the following shift 
in the momentum fraction $x$ for the parton inside the nucleus ($t$-channel):
\ben
x\to x\left(1+C_d \frac{\xi^2(A^{1/3}-1)}{-\hat t}\right),
\label{xshift}
\een
where $C_d=1$ ($C_A/C_F$) if parton $d$ is a quark (gluon) 
in the partonic channel $ab\to cd$. The reason for the values of 
$C_d$ is that $\xi^2$ has been  determined for coherent quark scattering~\cite{Qiu:2003vd}.
For the hard scattering partons that couple to the nucleus through  
other channels ($c$ in the $u$-channel and $a$ in the $s$-channel), 
similar shifts in the  momentum fraction $x$ have been derived in~\cite{Vitev:2006bi}. 
These can be obtained easily from Eq.~(\ref{xshift}) by substituting 
$\hat t \to \hat u$ and  $\hat t \to \hat s$, respectively, and by keeping 
track of the flavor of partons $c$ and $a$: $C_d \to C_c$ and   $C_d \to C_a$. 
In our paper we have taken into account the shift in $x$ consistently 
for all three  channels.

It is important to emphasize the similarities and differences between the dynamical shadowing 
calculation performed by Qiu and Vitev and the nuclear enhancement in the dijet (dihadron) 
imbalance presented in our last section. Qiu and Vitev studied the multiple scattering contributions 
to the dihadron differential cross section directly. These contributions are power 
suppressed by the hard  scale, for example $ \propto A^{1/3} \xi^2 / \hat t$  for the $t$-channel. 
In Refs.~\cite{ Qiu:2004da, Vitev:2006bi} the coherent multiple scatterings 
have been resummed and have been shown to lead to the shift in $x$ specified in Eq.~(\ref{xshift}). 
When this shift is small, only the first term in the resummed series is important and it is 
directly proportional to the twist-4 correlation function $T_{q,g/A}^{(I,F)}(x)$. 
On the other hand,  here we calculate the multiple scattering contribution to the 
$q_\perp^2$-weighted differential cross section. 
In the evaluation of the double scattering  contribution to this weighted cross section 
we can set $k_\perp=0$ in the hard parts. Because of the same partonic scattering origin, 
the dijet imbalance enhancement $\Delta\langle q_\perp^2\rangle$ and the nuclear modification 
factor $R_{dA}^{(2)}$ depend on the same characteristic scale $\xi^2$.

As the parton from the proton undergoes multiple scattering in the nucleus before 
the hard collisions,  it can lose energy due to medium-induced gluon bremsstrahlung. 
The spectrum of this initial-state energy loss was first derived in~\cite{Vitev:2007ve}.
At collider energies, for mid and forward rapidities ($y\geq 0$) 
even particles of small $p_\perp$ 
come from partons of very high energy in the rest frame of the nucleus (A is at large negative $y$).
In this regime, initial-state cold nuclear matter energy loss can noticeably 
affect the experimentally measured cross sections~\cite{Vitev:2006bi}.  There is renewed 
interest in constraining its magnitude through measurements with electromagnetic 
final states, such as the Drell-Yan production~\cite{Neufeld:2010dz,Xing:2011fb}.  Initial-state
energy loss has been implemented in the evaluation of hadronic and jet 
observables~\cite{Vitev:2006bi,Sharma:2009hn,Vitev:2009rd}. Such studies build 
upon early cold nuclear matter energy loss 
phenomenology~\cite{Vogt:1999dw,Gavin:1991qk,Johnson:2000ph,Arleo:2002ph}.

For a generic differential cross section, medium-induced radiative corrections factorize 
from the hard scattering process and enter as an integral convolution. 
For initial-state energy loss, a simple change of variables presents the effect 
as a rescaling of the momentum fraction of the incoming parton from the proton~\cite{Neufeld:2010dz}:    
\begin{eqnarray}
\frac{d \sigma}{dPS} &=& \int dx^\prime dx  \left[ \int d \epsilon \, P(\epsilon)  
f\left(\frac{x^\prime}{1-\epsilon} \right)  \right]  f(x)
 \, \frac{1}{2 x^\prime x s} \langle |{\mathcal M} (x^\prime P^\prime,x P)|^2 \rangle
 (2\pi)^4 \delta^4(p_i-p_f) \, , \quad
\label{chvb}
\end{eqnarray} 
where $\langle |{\mathcal M} (x^\prime P^\prime,x P)|^2 \rangle=H^U_{ab\to cd}(\hat s, \hat t, \hat u)$ defined in Sec.~II. For a single ($n_g=1$) emitted gluon $dN^g(\epsilon)/d\epsilon = P(\epsilon)$ is the 
probability distribution for fractional energy loss $\epsilon$. In general, $n_g \neq 1$
and the probability distributions $P_{q,g}(\epsilon)$
is constricted from $dN^g(\epsilon)/d\epsilon$ in the independent Poisson 
approximation,  $\epsilon=\sum_i \omega_i/E$. In the soft gluon approximation
medium-induced radiative corrections always factorize from the hard short-distance 
scattering~\cite{Vitev:2007ve,Ovanesyan:2011xy}. For arbitrary kinematics, 
this factorization is always exact for parent quarks~\cite{Ovanesyan:2011xy}. Just like in the 
vacuum, beyond the soft gluon approximation  medium-induced splitting kernels for parent 
gluons factorize only for polarization-averaged observables~\cite{Ovanesyan:2011kn}.
In this manuscript we work in the soft medium-induced splitting approximation 
which allows for an energy loss interpretation of the cross section suppression.
From Eq.~(\ref{chvb}), this effect is easily implemented through:
\begin{equation}
 f_{q,\bar{q}}(x')\to \int_0^1 d\epsilon\, P_{q}(\epsilon) f_{q,\bar{q}}\left(\frac{x'}{1-\epsilon}\right),
 \quad f_{g}(x') \to \int_0^1 d\epsilon\, P_{g}(\epsilon) f_{g}\left(\frac{x'}{1-\epsilon}\right).
\end{equation}

\bef
\psfig{file=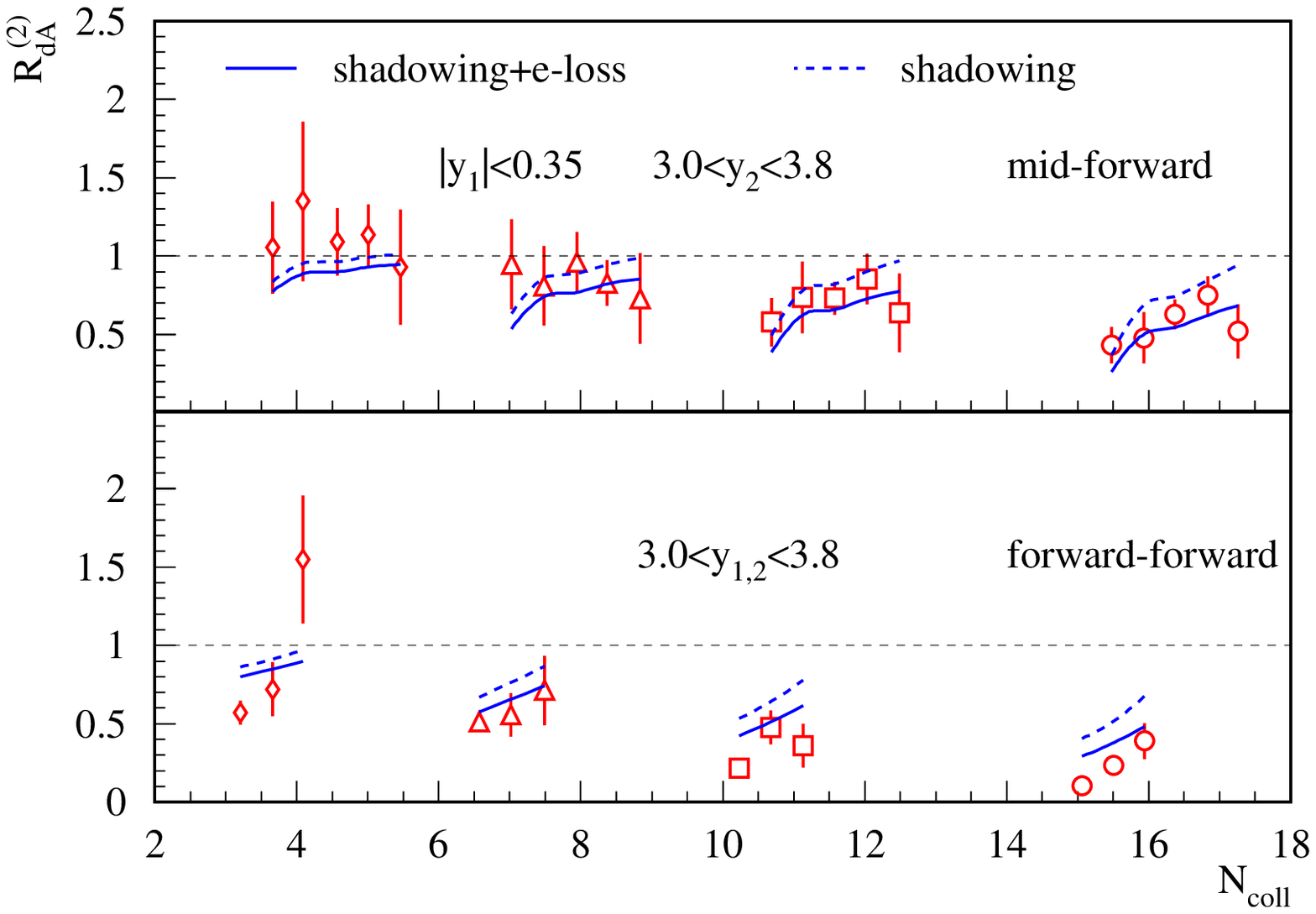, width=4.6in}
\caption{Nuclear modification factor $R_{dA}^{(2)}$ for back-to-back dihadron production 
in d+A collisions. Solid curves contain both dynamical shadowing and energy loss effect, 
whereas  dashed curves contain only dynamical shadowing. Both top and bottom panels have 
an associated particle in the forward-rapidity $3.0<y_2<3.8$ and $1.0<p_{\perp,\rm assoc}<1.5$ GeV 
region. From left to right there are four groups corresponding to a mean binary collision number 
$\langle N_{\rm coll}\rangle$=3.2, 6.6, 10.2, and 15.1, respectively, 
even though they have been offset from these actual values for visual clarity.
Top panel:  trigger particle at mid-rapidity $|y_1|<0.35$. Within each centrality selection, 
from left to right, the transverse momentum of the trigger particle is in the 
[1, 2], [2, 4], [2.5, 3], [3, 4], and [4, 7] GeV interval.
Bottom panel:
trigger particle at forward rapidity $3.0<y_1<3.8$. Within each centrality selection, 
from left to right, the transverse momentum of the trigger particle is in the 
[1.1, 1.6], [1.6, 2.0], and [2.0, 5.0] GeV interval. Data is from PHENIX~\cite{Adare:2011sc}.}
\label{RdA}
\eef
In Fig.~\ref{RdA}, we plot the nuclear modification factor $R_{dA}^{(2)}$ in d+Au 
collisions for dihadron production as a function of mean binary collision  
number $\langle N_{\rm coll}\rangle$ for both mid-forward (top panel) and forward-forward 
(bottom panel) correlated pairs. To take into account the centrality dependence, in 
both the transverse momentum imbalance and the dynamical shadowing  we have 
replaced $A^{1/3} \to A^{1/3} \langle N_{\rm coll}^{dA}(b) \rangle / 
\langle N_{\rm coll}^{dA}(b_{\rm min.bias}) \rangle$~\cite{Qiu:2004da}. 
In the calculation of initial-state radiative energy loss~\cite{Vitev:2007ve,Neufeld:2010dz} 
the average length of the medium is similarly scaled.
For both mid-forward and forward-forward cases, the associated particles are in the 
forward-rapidity region $3.0<y_2<3.8$ with momenta $1.0<p_{\perp,\rm assoc}<1.5$ GeV. 
For the mid-forward case the trigger particle is at mid-rapidity $|y_1|<0.35$, 
while for the forward-forward case the trigger particle is at forward rapidity $3.0<y_1<3.8$. 
The solid curves contain both dynamical shadowing and energy loss effects, 
whereas the dashed curves contain only the dynamical shadowing effect. As we can see, 
the calculated $R_{dA}^{(2)}$ give a very good description for the mid-forward 
correlated pairs.  The current experimental data still has large uncertainties, 
and is not able to further constrain the individual contribution of dynamic 
shadowing and the cold nuclear matter energy loss. For the forward-forward correlated pairs 
our formalism is still roughly consistent with the PHENIX measurements~\cite{Adare:2011sc}, 
though the agreements get slightly worse. In particular, for the most-central collisions 
the data seems to indicate a very strong suppression. Initial-state energy loss plays an 
important role in bringing the calculation closer to the data, as it does for inclusive
particle production at forward rapidities~\cite{Neufeld:2010dz}.
It will be interesting to see 
whether such  suppression factor can be explained within other formalisms, for example those 
based on color glass condensate~\cite{Albacete:2010pg, Stasto:2011ru}.

\subsection{Dihadron azimuthal correlations}
It is also  useful to present directly the dihadron azimuthal correlation distribution, 
see for example the STAR measurement in Fig.~\ref{angle}. In this case, the dihadron 
correlation can usually be approximated by two Gaussians for the near-side and the away-side, 
and a constant background:
\begin{equation}
CP(\Delta \phi)=\frac{1}{N_{\rm norm}}\frac{dN^{h_1h_2}}{d\Delta \phi}
\approx
B+\frac{A_N}{\sqrt{2\pi}\sigma_N}\exp\left\{-\frac{\Delta\phi^2}{2\sigma_N^2}\right\}
+\frac{A_F}{\sqrt{2\pi}\sigma_F}\exp\left\{-\frac{(\Delta\phi-\pi)^2}{2\sigma_F^2}\right\} \;.
\end{equation}
If we concentrate on the away-side peak, the area under the Gaussian peak $A_F$ 
is proportional to the production rate for approximately back-to-back hadron pairs. 
In going from p+p to d+Au collisions, we thus have: 
\ben
R_{dA}^{(2)}=\frac{A_F^{dAu}}{A_F^{pp}}.
\label{area}
\een
If one knows $A_F^{pp}$ for p+p collisions, from the theoretically calculated $R_{dA}^{(2)}$ in the 
last subsection, one  can predict the area under the Gaussian peak $A_F^{dAu}$ in 
d+Au collisions.

On the other hand, the width $\sigma_F$ of the away-side peak can be determined 
from the transverse momentum imbalance through Eq.~(\ref{sigmaF}). The change 
from p+p to d+Au will mainly depend on $\Delta\langle q_\perp^2\rangle$, which can 
be calculated in our formalism from Eq.~(\ref{imbalancehadron}). To obtain the dihadron 
azimuthal correlation in d+Au collisions, we first fit the dihadron correlation in p+p 
collisions to obtain $A_F^{pp}$ and $\sigma_F^{pp}$. We get:
\ben
B=0.0051, \qquad A_N=0.0122, \qquad \sigma_N=0.4238, \qquad A_F=0.0159, \qquad \sigma_F=0.7135.
\een
From Eq.~(\ref{sigmaF}), using $\langle p_{1\perp}\rangle\sim 2.54$ GeV 
and $\langle p_{2\perp}\rangle\sim 1.28$ GeV in p+p collision \cite{lesermes}, we find 
$\langle q_\perp^2\rangle_{pp}\sim 5.0$ GeV$^2$. We then calculate 
$\langle q_\perp^2\rangle_{dAu}=\langle q_\perp^2\rangle_{pp}+\Delta\langle q_\perp^2\rangle$ 
to obtain the away-side width $\sigma_F$ in d+Au collisions. On the other hand, 
with our calculation of $R_{dA}^{(2)}$ we get $A_F^{dAu}$ from Eq.~(\ref{area}). 
The dihadron azimuthal  correlations obtained this way are compared to the 
STAR experimental data~\cite{Braidot:2010ig} for both central and peripheral 
collisions in Fig.~\ref{angle}. The solid curves are calculated with $R_{dA}^{(2)}$ 
containing both dynamical shadowing and cold nuclear matter energy loss effects. 
The dashed curves are calculated with only dynamical shadowing. 
The constant offset is $B=0.01405$ for central collisions and $B= 0.0066$ for peripheral 
collisions. As we can see from 
the plot, our calculation gives a very good description of the experimental data 
in central collisions. For peripheral collisions, the agreements get worse. 
The main reason for the deviation comes from the fact that the experimental data for peripheral 
d+Au collisions show a clear broadening effect in the away-side width $\sigma_F$~\cite{Braidot:2010ig}. 
However, our calculated broadening 
$\Delta\langle q_\perp^2\rangle_{dAu}\propto A^{1/3} \langle N_{\rm coll}^{dA}(b) \rangle / 
\langle N_{\rm coll}^{dA}(b_{\rm min.bias})$ 
becomes quite small.

\bef
\psfig{file=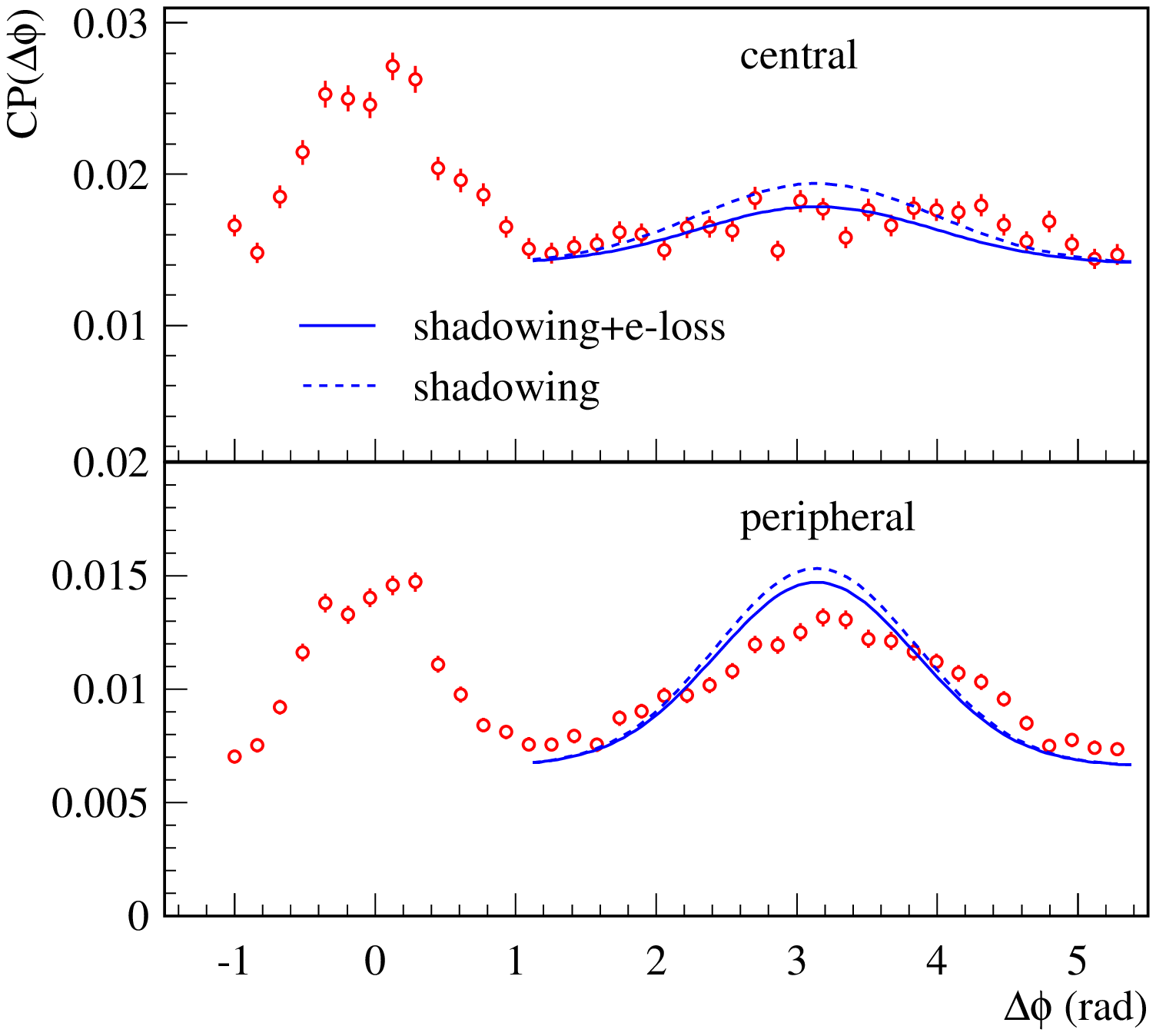, width=3.9in}
\caption{Azimuthal correlation associated with  back-to-back dihadron production in  
central (top) and peripheral (bottom) d+Au collisions. Theoretical curves are calculated for 
$\langle y_1\rangle=\langle y_2\rangle=3.2$ and $\langle p_{1\perp}\rangle=2.68$ GeV 
and $\langle p_{2\perp}\rangle=1.31$ GeV in d+Au collision~\cite{lesermes}. Data is from STAR \cite{Braidot:2010ig}.  }
\label{angle}
\eef

%%%%%%%%%%%%%%%%%%%%%%%%%%%%%
\section{Conclusions}
In summary, by taking into account both initial- and final-state multiple parton 
scattering inside the nucleus, we calculated in perturbative QCD the increase in the transverse 
momentum imbalance (nuclear-induced broadening) of dijet and dihadron production in high energy 
p+A (d+A) collisions relative to the more elementary p+p collisions. 
The nuclear-induced broadening can be used to calculate the width of the away-side peak 
in dihadron correlation measurements. For phenomenological applications, we combined   
our new theoretical findings with previously derived coherent power correction (dynamical shadowing) 
and cold nuclear matter energy loss results. Perturbative QCD calculations that take these 
effects into account were recently shown to give a good description of forward rapidity 
single inclusive particle production in d+Au collision at RHIC. In this manuscript we provided 
the corresponding evaluation for dihadron cross sections and correlations relevant to
the new STAR and PHENIX  measurements. With cold nuclear matter parameters constrained by data on 
deep inelastic scattering on nuclei, we found that the calculated nuclear modification factor 
is roughly consistent with the PHENIX  experimental data. Finally, by combining the calculated 
width of the away-side peak and the nuclear 
suppression factor, we were able to describe reasonably well the dihadron azimuthal correlations 
measured by the STAR experiment. Even though we need the baseline from p+p collisions, our formalism 
does describe the effects of cold nuclear matter in going from p+p to d+Au collisions pretty well 
for mid-mid,  mid-forward, and forward-forward correlated hadron pairs at RHIC.

%%%%%%%%%%%%%%%%%%%%%%%%%%%%%
\section*{Acknowledgments}
We thank L. Bland, E. Braidot, A. Ogawa, and X. Li for helpful correspondence 
on the STAR experimental data, and B. Meredith for the helpful correspondence on the 
PHENIX experimental data. This research is supported by the US Department of 
Energy, Office of Science, under Contract No.~DE-AC52-06NA25396 and DE-AC02-98CH10886, 
and in part by the LDRD program at LANL, NSFC of China under 
Projects Nos. 10825523 and 10875025.

%%%%%%%%%%%%%%%%%%%%%%%%%%%%%

\end{document}